\documentclass[%
 reprint,
 amsmath,amssymb,
 aps,
]{revtex4-1}

\usepackage{graphicx}% Include figure files
\usepackage{dcolumn}% Align table columns on decimal point
\usepackage{bm}% bold math
\begin{document}

\preprint{APS/123-QED}

\title{Ising spin model using Spin-Hall Effect (SHE) induced \\magnetization reversal in Magnetic-Tunnel-Junction}% Force line breaks with \\

\author{Yong Shim}
%  \altaffiliation[Also at ]{Physics Department, XYZ University.}%Lines break automatically or can be forced with \\
  \email{shim13@purdue.edu}
\author{Akhilesh Jaiswal}%
%  \email{Second.Author@institution.edu}
\author{Kaushik Roy}%
%  \email{Second.Author@institution.edu}
\affiliation{%
 School of Electrical and Computer Engineering, Purdue University, West Lafayette, Indiana 47907, USA\\
%  This line break forced with \textbackslash\textbackslash
}%

\begin{abstract}
Ising spin model is considered as an efficient computing method to solve combinatorial optimization problems based on its natural tendency of convergence towards low energy state. The underlying basic functions facilitating the Ising model can be categorized into two parts, ``Annealing and Majority vote''. 
In this paper, we propose an Ising cell based on Spin Hall Effect (SHE) induced magnetization switching in a Magnetic Tunnel Junction (MTJ).
The stochasticity of our proposed Ising cell based on SHE induced MTJ switching, can implement the natural annealing process by preventing the system from being stuck in solutions with local minima. 
Further, by controlling the current through the Heavy-Metal (HM) underlying the MTJ, we can mimic the majority vote function which determines the next state of the individual spins.
By solving coupled \textit{Landau-Lifshitz-Gilbert} (LLG) equations, we demonstrate that our Ising cell can be replicated to map certain combinatorial problems.
We present results for two representative problems - Maximum-cut and Graph coloring - to illustrate the feasibility of the proposed device-circuit configuration in solving combinatorial problems. Our proposed solution using a Heavy Metal (HM) based MTJ device can be exploited to implement compact, fast, and energy efficient Ising spin model. 
\end{abstract}

\pacs{Valid PACS appear here}% PACS, the Physics and Astronomy
\maketitle

%\tableofcontents

\section{introduction}

Efficient computing models for combinatorial optimization problems have attracted considerable research interest. This trend is in sync with the flood of information in the present Internet-of-Things (IoT) era. Such huge amount of collected data from multiple sensors is required to be handled properly for certain purposes. As an example, real time management of a smart building (including lighting, cooling and heating) requires complex optimization using data from multiple sensors \cite{baz_iot}. However, solving optimization problems in an efficient way based on conventional computing model is very challenging. Typically, to find an optimum solution for such problems, a performance index has to be computed and compared for every possible input combinations. \cite{yamao_isingchip2} However, the computational cost associated with a combinatorial optimization problem, increases exponentially with the number of variables. Moreover, if we consider the process of problem solving based on conventional computing (by following a sequential fetch, decode, and execute cycles), finding an optimum (even near optimum) solution seems infeasible keeping in view the energy and power requirements. 

Ising model has been researched extensively owing to its simple architecture and inherent ability to solve combinatorial optimization problems \cite{barah_ising,cipra_ising}. Recently, an Ising model based on a nano-magnet with a HM, in the 'telegraphic' switching regime, was proposed in \cite{datta_ising}. In the present manuscript, we propose an Ising cell based on controlled stochastic switching dynamics of the magnetization direction in a nanomagnet with an underlying HM layer. A numerical simulation framework based on the stochastic \textit{Landau-Lifshitz-Gilbert} (LLG) equation is developed to analyze the switching characteristic of the proposed device. Further, by solving coupled stochastic LLG equations and SPICE simulations, we show solutions for some representative combinatorial problems obtained by using our proposed Ising cell.

\begin{figure}
\includegraphics[width=3.4in]{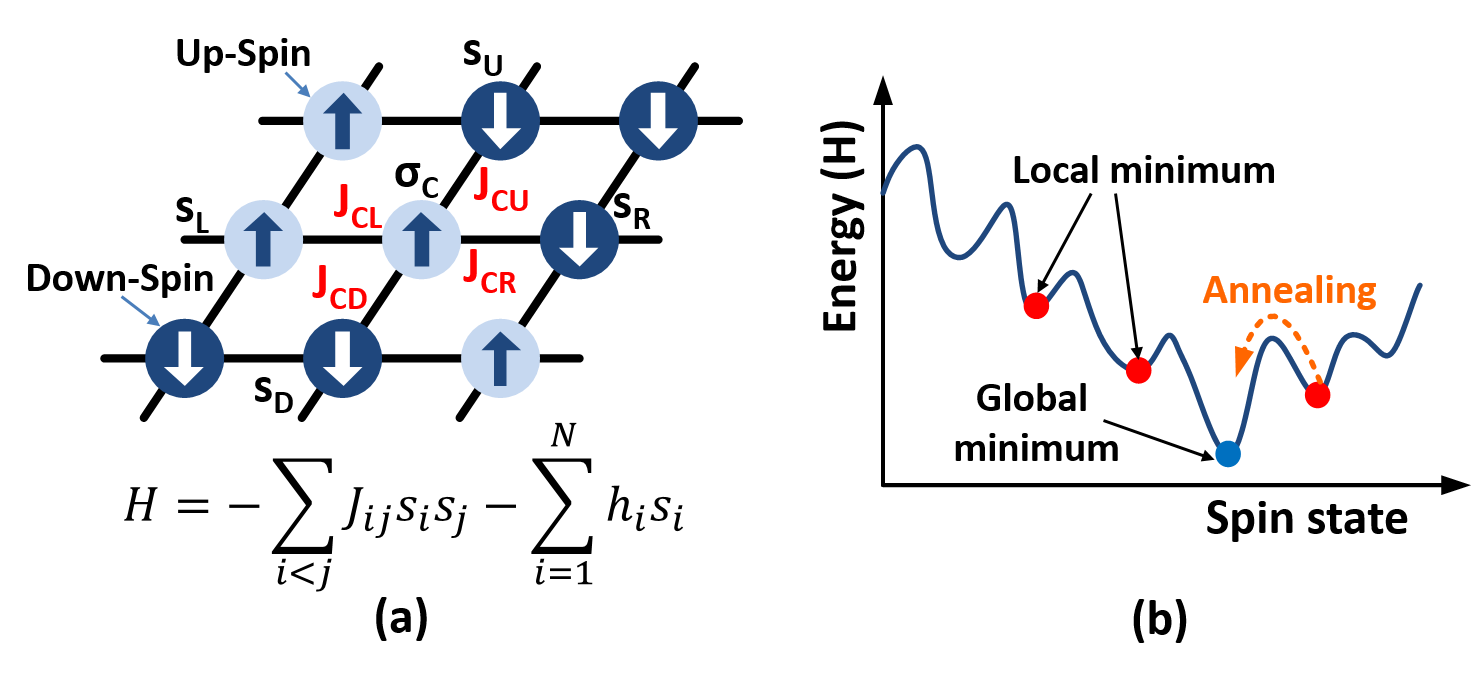}% Here is how to import EPS art
\caption{(a) Conventional Ising spin model consists of spin(s$_{i}$), interconnection weights (J$_{ij}$), and external magnetic field (h$_{i}$). Definition of Hamiltonian ($H$), total energy of the system, is also shown in below (b) Energy of the system changes depending on the states of the spins. The energy profile has a global minimum energy state and multiple local minimum energy states. The annealing process prevent the system being stuck into a local minima. }
\label{fig1}
\end{figure}

Before we describe the proposed device, we would give a brief introduction of the Ising spin model in general. 
The Ising model considers the behavior of magnetic spins and the coupling between them. Fig. \ref{fig1}(a) illustrates a simple view of the Ising model and the definition of associated Hamiltonian ($H$) - the total energy of the system. The model consists of individual spin state (s$_{i}$), interconnection coefficient between two spins (J$_{ij}$), and external magnetic field (h$_{i}$). Each spin can have one of the two states, up and down, and there are four interconnections with neighbors in this model. The spin at the center is named as s$_{C}$ and the four neighbors are s$_{U}$, s$_{D}$, s$_{L}$, and s$_{R}$. The interconnection weights between s$_{C}$ and its neighbors are denoted as J$_{CU}$, J$_{CD}$, J$_{CL}$, and J$_{CR}$, respectively. These weights model the coupling between spins and are used to determine next state of the spins. For example, if J$_{CU}$ has positive sign (i.e. +1), it implies s$_{U}$ tries to align s$_{C}$ parallel to itself. Likewise, a neighboring spin with a negative weight tries to align the given spin anti-parallel to itself. Since there are four neighbors, the next state of s$_{C}$ is decided based on a majority vote - if majority of the neighbors of a given spin state want to keep the given spin in +1 direction, then the next state of s$_C$ will be +1, else it would be -1. 
The Hamiltonian (H) also changes as the states of the spins are updated. % changed
Hence, once the problem is mapped to the system properly (by programming the weights for each interconnection), the system tries to reach the energy minimum state by switching the states of the spins through the aforementioned majority coupling. When the system reaches the global minimum energy state, the solution is obtained by examining the final states of the spins. 

Fig. \ref{fig1}(b) shows the total energy ($H$) of the system as a function of different spin states. As shown in the figure, the energy profile has a global minimum, and also multiple local minimum states. This implies that the system could easily get stuck at the local minimum state during the process of problem solving (\textit{i.e} the system evolves to different states through coupling). To avoid the system being stuck into a local minima, annealing process, such as ``Simulated Annealing (SA)'' \cite{kirkp_sa,huse_sa} and ``Quantum Annealing (QA)'' \cite{finni_qa,kadow_qa,santo_qa} has been proposed. During a SA process in conventional Ising model, the system starts from a known initial states at a non-zero temperature. The system then evolves towards the minimum state of the Hamiltonian by lowering its temperature gradually. In contrast, in a QA, the temperature can be replaced by a quantum mechanical effect, such as probabilistic quantum tunneling \cite{kadow_qa}. Whether it is QA or SA, annealing always includes some kind of randomization of the next state logic, to get the system out of the local minima.

Despite the fact that SA and QA can find the lowest energy state of the Ising spin model efficiently, the implementation of such a system needs control of the state of each spin and coupling between them. Also, the state of individual spins needs to be monitored for total energy of the system which is challenging from hardware perspective. This is why hardware implementation of the Ising model did not receive much attention, even though theoretical background has been widely explored by the research community. 
Recently, hardware implementations of the Ising spin model have been proposed using Superconducting material \cite{johnson_qa} and CMOS only implementation \cite{yamao_isingchip}. % Not clear
In CMOS circuits, the annealing process can be implemented by generating a random address that is used to choose a specific spin to be flipped \cite{yamao_isingchip}.
However, such implementations require complex hardware for random number generation, address decoding, and write operation for the specific spin state. These series of operations have to be executed multiple times to get the system state out of the local minima. Furthermore, randomizing spin states based on random number generation can potentially move the system to totally different state so that the system might not reach the global minimum state eventually. 

Another difficulty in implementation of the Ising model is due to the complexity of the majority voting circuit. As explained, the next state of each spin is determined by the interactions with all neighboring states. 
Multiple solutions might be possible to implement majority function based on digital, analog, and even with mixed-mode design. However, all of these implementations are expensive in terms of silicon area and power consumption due to its multi-input nature and complexity of operation.

The motivation of our research arises from the fact that a spintronic device like a Heavy-Metal (HM) based Magnetic-Tunnel-Junction (MTJ) can potentially provide the aforementioned two important characteristics required for the Ising model \textit{viz.} 1) stochasticity (random spin flip required for annealing) 2) majority voting.
In order to mimic the aforementioned functionality, a nanomagnet is required that is able to switch its states with a certain probability to facilitate the annealing process and also change its state based on a majority vote which is crucial for the system to evolve towards minimum energy state. In the next section, we describe the mapping of the magnetization dynamics of a nanomagnet to the Ising operations such as a natural randomizer and a majority vote logic. 

\section{From device to Ising model}
\begin{figure}
\includegraphics[width=3.4in]{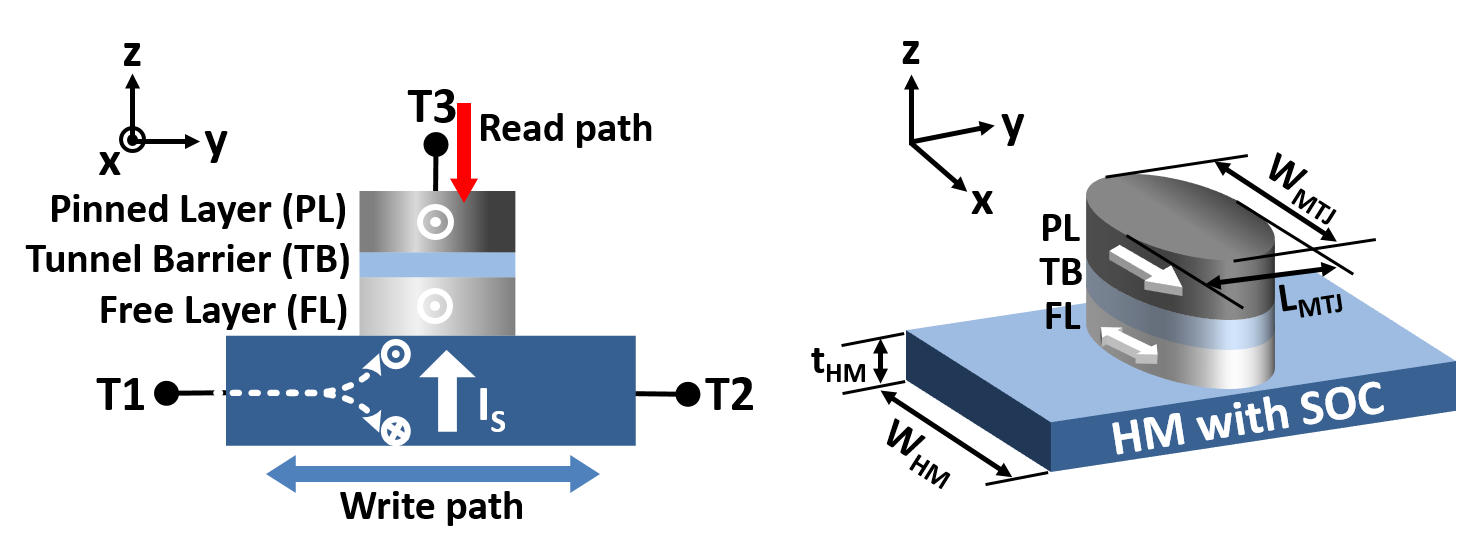}% Here is how to import EPS art
\caption{3-terminal SHE-MTJ device with MTJ on the top of the Heavy-Metal (HM) layer. The magnetization of FL with in-plane magnetic anisotropy (IMA) can be manipulated by the current flow through the HM layer. The decoupled read/write path can provide design flexibility.}
\label{fig2}
\end{figure}

Let us first describe the basic device structure used for our Ising model and its principle of operation. Fig. \ref{fig2} shows a three terminal device structure, consisting of the conventional MTJ stack formed by a Tunneling Barrier (TB) sandwiched between two nanomagnets. Since the magnetization of the upper ferromagnetic layer is fixed, it is termed as the Pinned Layer (PL). On the other hand, the magnetization of the bottom layer, denoted as the Free Layer (FL), can be manipulated by an incoming spin current. Depending on the direction of the FL, the MTJ structure can have two stable states. If the magnetizations of the two ferromagnetic layers are in the same direction, it is in the parallel configuration (P), otherwise it is in the anti-parallel configuration (AP). These two states have different resistances across the vertical direction of the device. Typically, AP state has a higher resistance and the ratio between P and AP states is defined as the Tunnel MagnetoResistance (TMR) ratio.

The FL is in contact with a heavy metal (HM) like Pt or Ta. On passing a charge current through the HM, from terminal T1 to T2, the direction of magnetization in the FL becomes parallel to that of the PL. When the direction of charge current flowing through the HM is reversed (from T2 to T1), the FL switches its direction such that it is now anti-parallel to the PL. The switching of FL due to a charge current flowing through the HM can be attributed to the large spin-orbit-coupling (SOC) of the HM. SOC is a relativistic effect that theoretically originates from the full relativistic wave-equation. In the current scenario, due to SOC based effects like the Spin Hall Effect (SHE), the electrons flowing through the HM are deflected such that up and down spins split. As shown in Fig. \ref{fig2}, the up-spins are deflected in the +z direction and down-spins in the -z direction. This spin splitting results in a resultant spin current flowing in the +z or -z direction, based on the direction of the charge current. The spin current ($I_s$), thus generated due to the charge current flowing in the HM, exerts a torque on the magnetization direction of the FL, making it parallel or anti-parallel to the PL. Thus, a charge current flowing between the terminals T1 and T2 can switch the state of the MTJ from $R_{AP}$ to $R_P$ or \textit{vice-versa}.
In order to sense the resistance of the MTJ, a voltage can be applied between terminals T1 and T3/T2. By sensing the current flowing through the MTJ stack (or detecting the voltage level across the device) one can conclude if the current state of the MTJ is $R_{AP}$ or $R_P$. Based on the above description, the three terminal structure of the device shown in Fig. \ref{fig2} offers the following desirable characteristics 1) the write and read path are decoupled and can be independently optimized 2) the efficiency of spin current generated from the charge current flowing through the HM, can be greater than 100\% \cite{intel_she}, resulting in low write-energy 3) by controlling the amount of current flowing through the HM, one can not only alter the switching probability but also implement a majority function. Later in the manuscript, we would describe how these desirable characteristics of the HM based MTJ device can be used efficiently to mimic the various operations required for the Ising spin model.

We would now describe the equations used for modeling the device shown in Fig. \ref{fig2}. Under an external excitation, for example a magnetic field or a spin current, the magnetization dynamics of the FL can be obtained by the well known \textit{Landau-Lifshitz-Gilbert} (LLG) equation with an additional term for the spin transfer torque proposed by Slonczewski \cite{slonc_spin}, 
\begin{eqnarray}
\frac{d\widehat{m}}{dt}=-\gamma (\widehat{m} \times H_{eff}) + \alpha (\widehat{m} \times \frac{d\widehat{m}}{dt})\nonumber \\
+ \frac{1}{q{N_{s}}}(\widehat{m} \times I_{s} \times \widehat{m})
\end{eqnarray}
\lowercase{w}here $\hat{m}$ ̂is the unit vector corresponding to the direction of magnetization in the FL, $\gamma = 2\mu_{B}\mu_{0}/\hbar$ is the gyromagnetic ratio for electron, $\alpha$ is Gilbert damping ratio, $H_{eff}$ is the effective magnetic field including the shape anisotropy field \cite{beleg_demag} and uniaxial interface anisotropy field \cite{scaling_aniso}, $N_{s}=M_{s}V/\mu_{B}$ is the number of spins in the FL volume $V$ ($M_{s}$ is saturation magnetization and $\mu_{B}$ is Bohr magneton), $I_{s}$ in equation (1) is the spin current flowing through the FL in the +z or -z direction.

Based on recent experimental studies \cite{hirsc_she,pai_she,miron_she,liu_she,liu_she2}, the spin current generated due to a charge current flowing through the HM can be estimated as
$I_{s}=\theta _{SH}\frac{W_{MTJ}}{t_{HM}}I_{Q}$, where $I_{Q}$ is the charge current flowing through the HM, $\theta_{SH}$ is the spin-hall angle \cite{pai_she}, $W_{MTJ}$ and $t_{HM}$ are device dimension parameters, shown in Fig. 2. The details of the device parameters we used for benchmarking are summarized in Table I.
\begin{figure}
\includegraphics[width=3.4in]{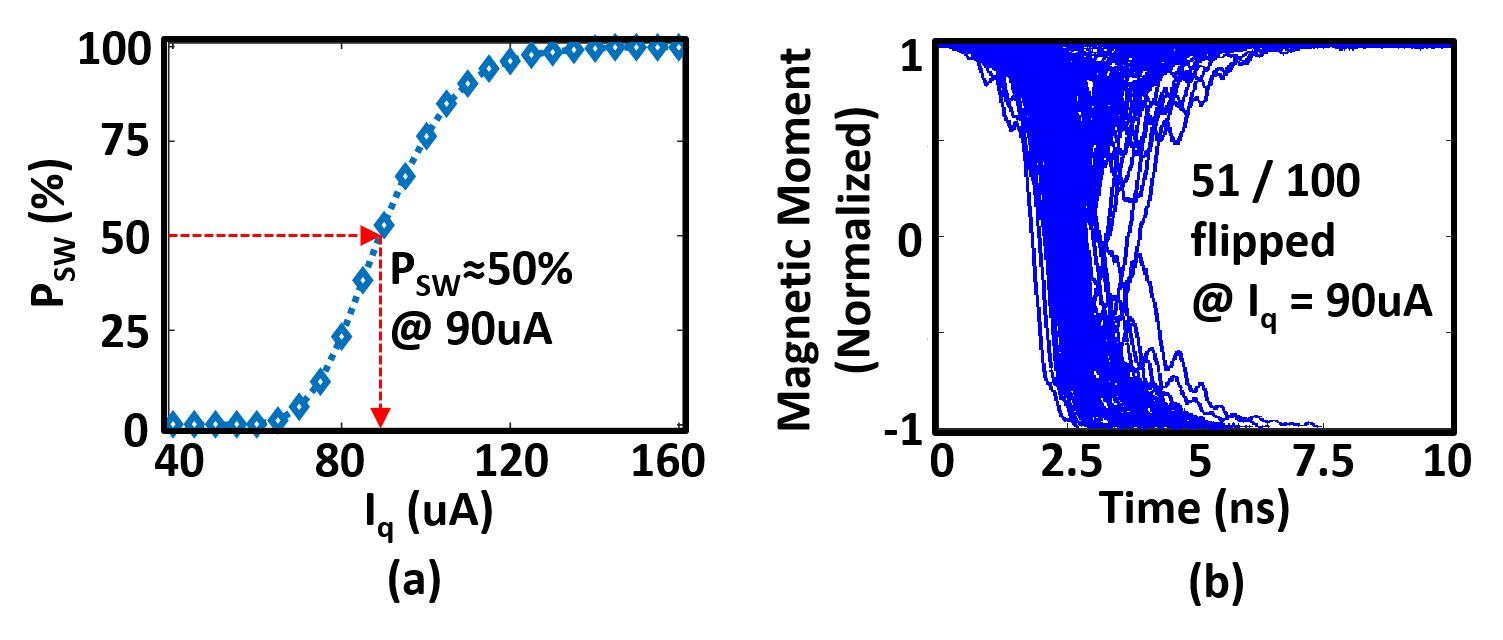}% Here is how to import EPS art
\caption{(a) Switching probability (P$_{SW}$) versus input charge current through the HM layer (I$_{q}$). The current pulse with different height (40 $uA$ to 160 $uA$, 5 $uA$ step) is applied for t$_{WRITE}$ (=3 ns) and then wait for t$_{RELAX}$ (=6 ns) before check the final magnetization state. (b) Magnetic moment change (normalized) from 1 to -1 during 100 write operation with I$_{q}$= 90 $uA$ for 3 ns. Total 51 out of 100 cases are flipped as can be expected from Fig. \ref{fig3}(a)}
\label{fig3}
\end{figure}
\begin{table}[b]%The best place to locate the table environment is directly after its first reference in text
\caption{\label{tab:table1}%
Device simulation parameters
}
\begin{ruledtabular}
\begin{tabular}{lcdr}
\textrm{Parameters}&
\textrm{Value}\\
% \multicolumn{1}{c}{\textrm{Decimal}}&
% \textrm{Right}\\
\colrule
Free layer area & ($\pi/4$) $\times$ 45 $\times$ 112.5 nm$^{2}$ \\
Free layer thickness & 1.5 nm \\
Heavy-metal thickness, $t_{HM}$ & 2.3 nm \\
Saturation magnetization $M_{s}$ & 1257.3 emu/cm$^3$ \cite{ikeda_pma} \\
Spin-Hall Angle, $\theta_{SH}$ & 0.3 \cite{intel_she} \\
Gilbert damping factor $\alpha$ & 0.1 \\
Energy barrier $E_{B}$ & 60 $KT$ \\
MgO Thickness, $t_{MgO}$ & 1.4 nm \\
MTJ resistance, $R_{P}$($R_{AP}$) & 8.56 (18.31) $K$\text{$\Omega$} \\
Resistivity of HM, $\rho_{HM}$ & 200 $\mu \Omega$ - cm \cite{intel_she}\\
Pulse width $t_{PW}$ & 3 ns \\
Temperature $T_{K}$ & 300 $K$ \\
Supply Voltage, $V_{DD}$ & 1 $V$ \\
\end{tabular}
\end{ruledtabular}
\end{table}

Finally, to model the effect of thermal noise at non-zero temperature, the thermal noise is accounted as a random thermal field \cite{schol_htherm}, $H_{thermal}=\sqrt{\frac{\alpha }{1+\alpha ^{2}}\frac{2K_{B}T}{\gamma \mu _{0}M_{s}V\delta _{t}}G_{0,1}}$, where $G_{0,1}$ is a Gaussian distribution (zero mean, unit standard deviation), $K_{B}$ is the Boltzmann constant, $T$ is the temperature and $\delta _{t}$ is the time step. Under the effect of thermal noise, the switching behavior of the MTJ during the write operation due to the charge current flowing through the HM layer, becomes stochastic in nature. Also, the probability of switching changes according to the magnitude of the input charge current. Fig. \ref{fig3}(a) illustrates a graph showing switching probability (P$_{SW}$) versus input charge current (I$_{q}$) through the HM layer. The charge current pulse was applied for a fixed time (3 ns) and an additional 6 ns wait time was included for the magnetization to relax. The amount of charge current varies from 40 $uA$ to 160 $uA$ with 5 $uA$ step, and 10$^{4}$ simulations were executed for each simulation step to get a switching probability. As shown in the figure, when an input current of 90 $uA$ is applied for 3 ns, the P$_{SW}$ becomes roughly 50 \%. Fig. \ref{fig3}(b) shows the normalized magnetic moment change with 100 write operations (assuming that the initial magnetization is 1 and is being flipped to -1 direction) when the input charge current is kept at 90 $uA$ for 3 ns followed by 6 ns of relaxation time. It can be observed, 51 out of 100 cases  flipped which is close to the ratio one would expect from Fig. \ref{fig3}(a). 

This stochastic flipping nature of the nanomagnet can be potentially exploited as a natural randomizer - one of the key elements for the `natural annealing' process. The baseline idea of a general annealing process in Ising model lies in perturbing the spin states randomly to get the system out of the local minimum energy state. 
Thus, while switching the state of the MTJ, we can tune the input charge current flowing through the HM such that the MTJ switches with the desired probability.
The write current can be controlled with ease by adopting simple CMOS peripherals which will be explained later. 
It is worth noting here that the proposed natural annealing can also provide time-varying switching probability (by adjusting the input current value) which mimics the natural property of annealing through the temperature control. This helps to find a global minimum quickly when the system reaches the end of iterations.

\begin{figure}
\includegraphics[width=3.4in]{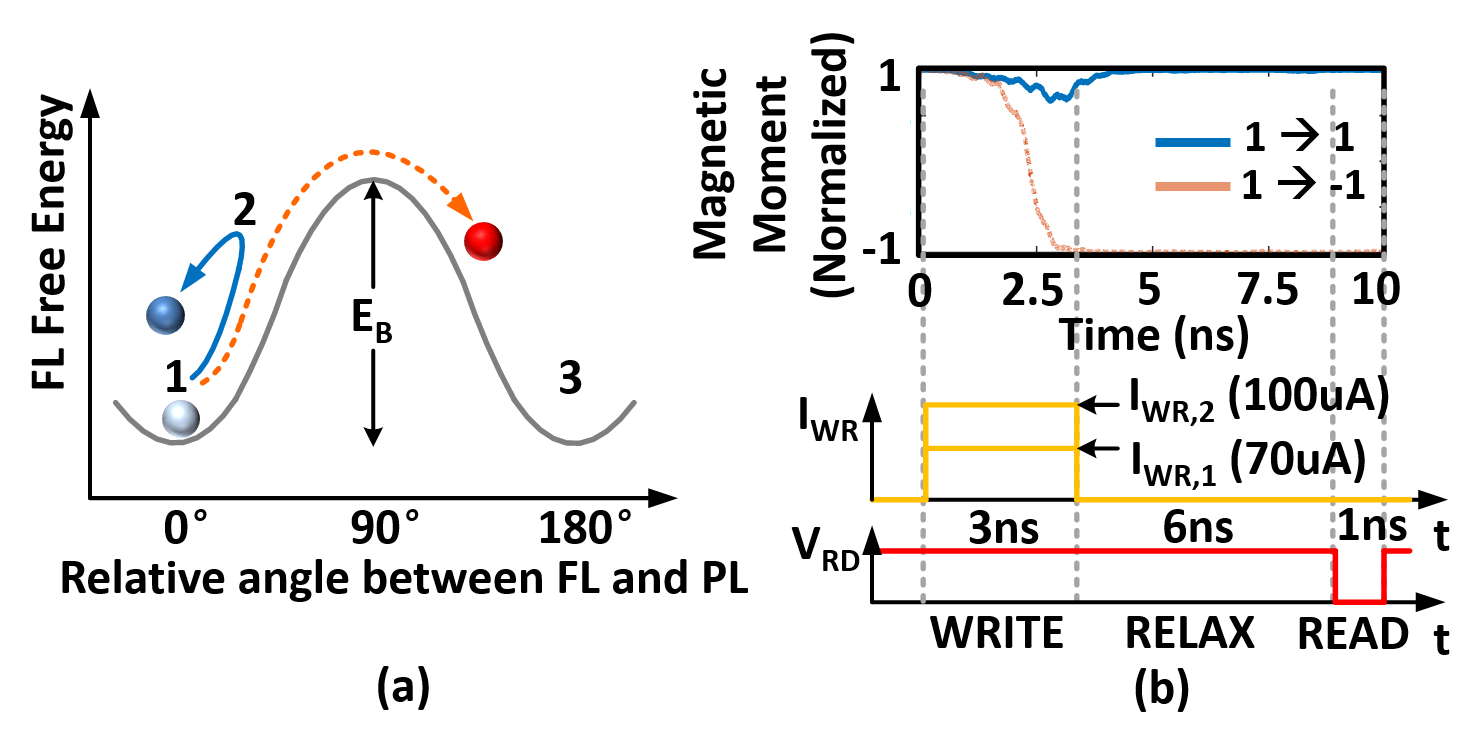}% Here is how to import EPS art
\caption{(a) Different switching behaviors depending on the magnitude of input charge current can be explained with energy profile of the FL of the MTJ device. (b) Timing diagram for two write operations and corresponding magnetic moment change. To exhibit successful and unsuccessful write operations, current pulses with different levels are presented to the HM layer with the timing in the figure.}
\label{fig4}
\end{figure}

Additionally, we can exploit another benefit from the nanomagnet due to the strong dependence of the switching process on the input charge current flowing through the HM. As explained, charge current flowing through the HM layer induces spin current in transverse direction at the FL-HM interface, which flips magnetization of the FL. The dependence of the flipping process on the input current can be explained by the energy profile of the FL. Assume that the angle between FL magnetization and the PL magnetization be represented by $\theta$. The FL energy as a function of $\theta$ is shown in Fig. 4(a), where the two stable states ($\theta$=0$^{\circ}$ and $\theta$=180$^{\circ}$) are separated by the energy barrier E$_{B}$. Here we assume that the MTJ changes its state from P ($\theta$=0$^{\circ}$) to AP ($\theta$=180$^{\circ}$) state. Once the input charge current is presented to the HM layer for a duration t$_{WRITE}$, spin current is induced based on the spin-hall effect (SHE) and makes the spin at point 1 to move uphill along the energy profile. If the charge current is not enough to move the spin across the energy barrier, the spin stops at the metastable state (point 2) and falls down again to the point 1 during t$_{RELAX}$. On the other hand, once enough charge current is applied, ultimately the spin will overcome the energy barrier and move to the point 3. Fig. \ref{fig4}(b) shows these situations using magnetic moment change and two current pulses with different magnitudes. 
Note that, in presence of thermal noise, a hard switching threshold current does not exists. A particular amount of current can only flip a magnet with certain probability unless the applied current is large enough to deterministically switch the magnet. 
% up to here, Akilesh

The decoupled write operation in HM based MTJ can be used as an efficient way to construct the majority vote function required for the Ising model. Let us consider a given connection between a particular spin state and one of its neighbors. We would represent the connection as (+1,+1), where the first number denotes the spin state (+1 represents up spin and -1 down spin), while the second number in the bracket represents the associated weight of the given spin and its neighboring spin.
As mentioned in the previous section, the next state of each spin is determined through the interaction with all connected neighboring spins. For example, if a neighbor has a state (+1,+1) or (-1,-1), then it would want the next state of the spin to be +1 (obtained by product of spin state and associated weight).
In case of (+1, -1) or (-1, +1), the neighbors would tend to make the next state of the given spin under consideration as -1. In the presence of multiple neighbors, a majority vote is taken to determine the next state of the spin under consideration. The hardware implementation of such a multiplication (product of spin state and associated weight) and majority vote functionality requires complex circuit. 
\begin{figure}
\includegraphics[width=3.4in]{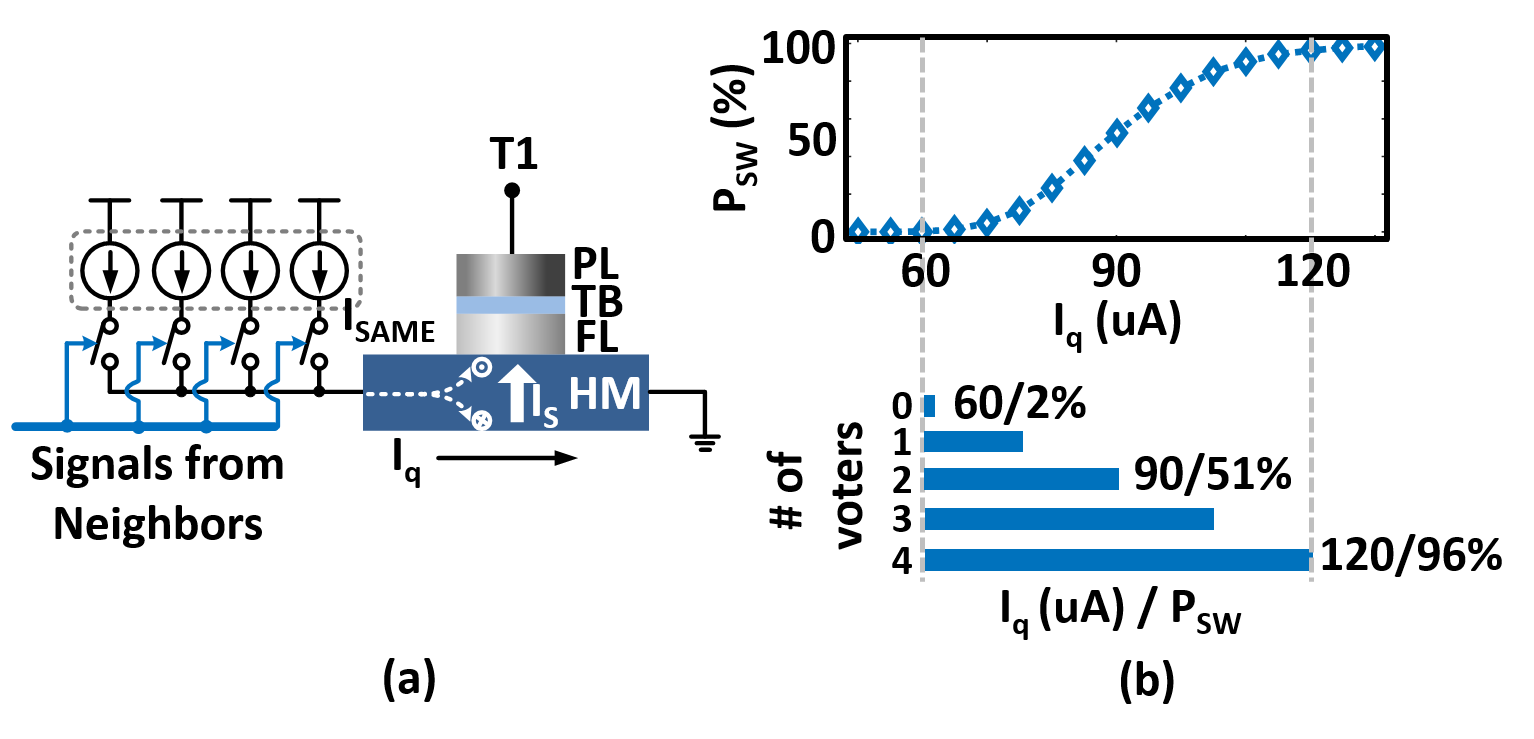}% Here is how to import EPS art
\caption{(a) Majority vote function is mapped to current dependent switching probability (P$_{SW}$) changes based on MTJ on the HM layer. To control the amount of current, multiple current sources with corresponding switches are used. (b) Inputs from neighbors are used to control the amount of current through the HM layer, which determines P$_{SW}$  }
\label{fig5}
\end{figure}

Our proposed HM based MTJ circuit that can efficiently implement the majority vote functionality is shown in \ref{fig5}(a).
The HM layer receives the input current proportional to the number of voters from the neighboring spin states using multiple current sources and switches. The corresponding P$_{SW}$ is depicted on the top of Fig. \ref{fig5}(b). Here we assume the possible current range for write operation is limited within 60 $uA$ to 120 $uA$. This range is equally divided among its neighbors. For instance, if there are four neighbors, each neighbor would contribute 15 $uA$ of write current by voting to  one of the two potential states. Based on this, if there are 0 voters, the current during the write operation becomes 60 $uA$ which leads to 2 \% P$_{SW}$. If there are 4 voters, then the current becomes 120 $uA$ and corresponding P$_{SW}$ becomes 96 \%. This directly mimics the general rule of majority vote - low P$_{SW}$ with less voters, high P$_{SW}$ with more voters. Note that, due to its probabilistic nature there are chances of spin flip into unwanted state. This is not an issue, it rather mimics the natural annealing process as discussed earlier.
\begin{figure}
\includegraphics[width=3.4in]{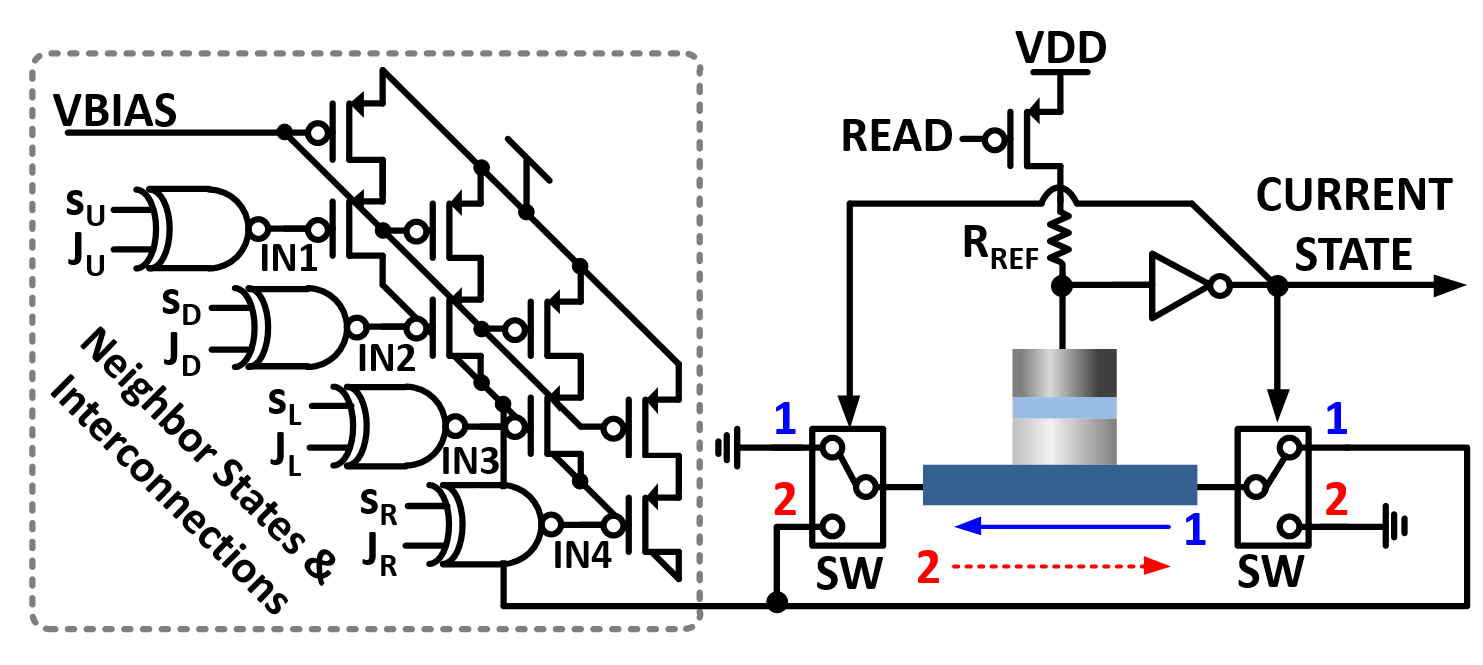}% Here is how to import EPS art
\caption{The proposed device-circuit configuration for single Ising spin model. To convert current state of the spin to digital value, series of transistor and reference resistor (R$_{REF}$) are used with output inverter. For the majority function and write operation, current source with CMOS logic gates are adopted. Inputs from neighbors are used to control the amount of charge current, which determine the switching probability (P$_{sw}$)}
\label{fig6}
\end{figure}

The overall device-circuit configuration for single spin model is shown in Fig. \ref{fig6}. For read operation, reference resistor (R$_{REF}$) and the switch transistor are used in series on the top of the nanomagnet. The resistance of R$_{REF}$ is in between two stable resistance states of the MTJ device. Hence, if the MTJ resistance is smaller than R$_{REF}$ (in parallel magnetization configuration), the output of the inverter becomes high, and vice versa. 
The `current state' signal is then used to choose the direction of the current flowing through the HM.
For the majority vote and write operation, there are current sources with multiple branches (each with stacked transistors, one for biasing and the other for switch operation) and XNOR gates to combine information from the neighbors (product of spin state and interconnection weights). Note, here we assumed that there are four neighbors. The number of neighbors can be extended by adding more branches on current source and controlling the amount of current from each branch. Thus, 1) the XNOR gate (which implements the multiplication function), 2) the transistor switches and the dependence of P$_{sw}$ on input current (which mimics the majority vote logic), 3) the `current state' logic which controls the direction of current flow and 4) the inherent stochasticity of switching (representing the annealing process) constitute our basic Ising cell, that can be replicated to implement combinatorial optimization problems.

\section{Problem Mapping and Results}

\begin{figure}
\includegraphics[width=3.4in]{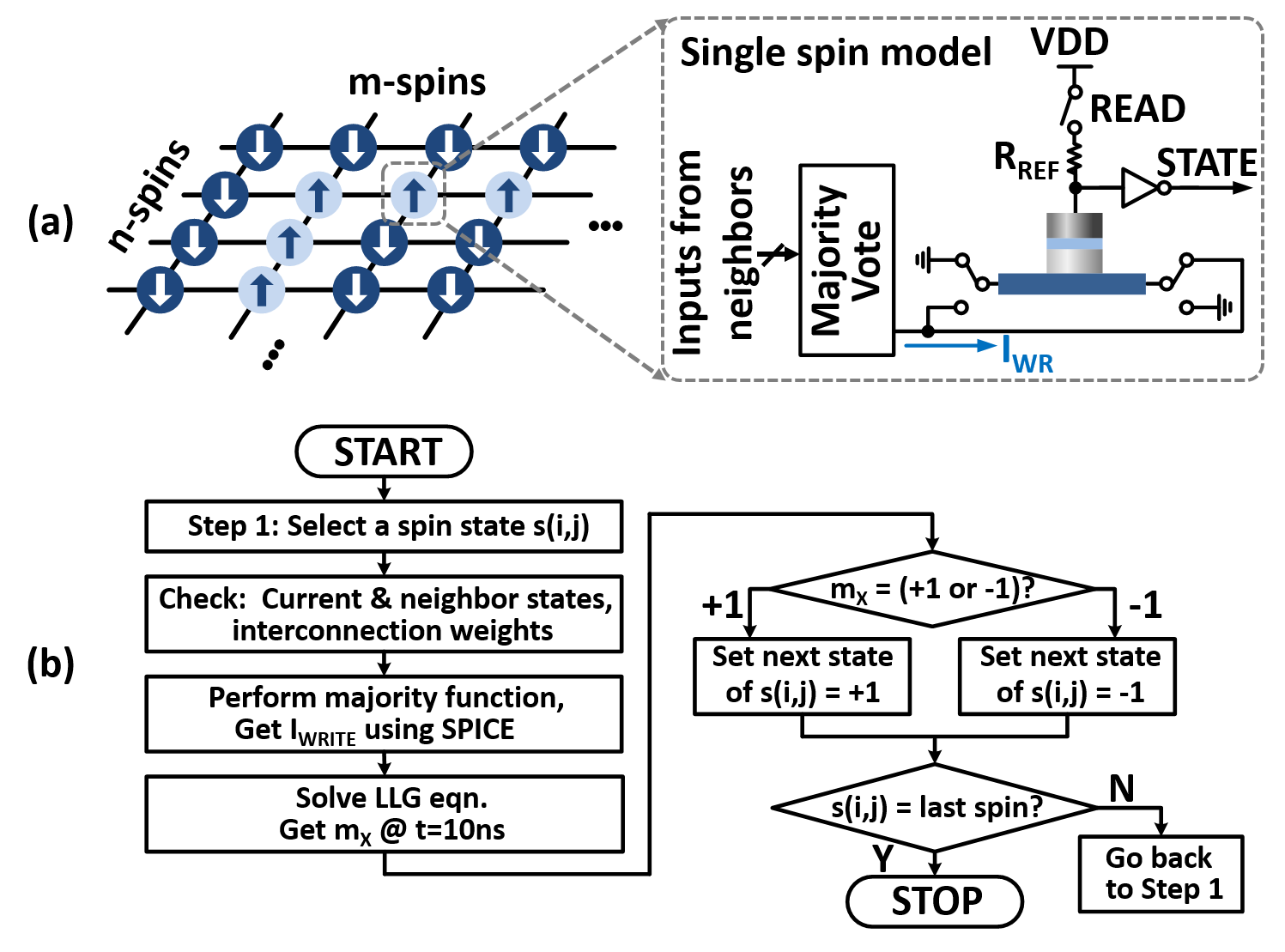}% Here is how to import EPS art
\caption{(a) Ising spin model based on the proposed hardware implementation for a single spin state. (b) The process of spin states update. The majority function based on neighboring states and interconnection weights is processed using SPICE simulation. The devices characteristics are studied by numerically solving a LLG equation.}
\label{fig7}
\end{figure}

In this section, we would first elaborate how the basic Ising cell described in the previous section, can be used to map combinatorial problems. For the Ising spin model shown in Fig. \ref{fig7}(a), each spin state consists of an MTJ device with underlying HM layer and associated interface circuits. The interconnection weights between neighboring spins are a function of the specific problem to be solved. These weights are programmed into the system before the Ising model tries to converge towards the minimum energy state. Based on the initial spin states and associated weights, the system evolves by updating the state of each spin through the coupling (annealing and majority vote) described earlier in the manuscript. The time evolution of the Ising system in our simulation framework was achieved by following the steps shown in  Fig. \ref{fig7}(b). After selecting a specific spin to be updated, the neighboring states and the weights associated with each interconnection are examined. Then, the number of voters for the potential next state (either +1 or -1) and corresponding charge current for write operation is obtained through SPICE simulations.  The resultant current level is then fed to the LLG solver to analyze the magnetization switching behavior of the HM based MTJ. After applying the current pulse for a duration of 3 ns to the HM layer followed by 6 ns of relaxation time, the final direction of the FL magnetization is examined which determines the next state of the spin under consideration. This process is repeated until the state of the all the spins in the system is updated. Then, the system computes the Hamiltonian of the next state to check whether the system has found a solution. 

\begin{figure}
\includegraphics[width=3.4in]{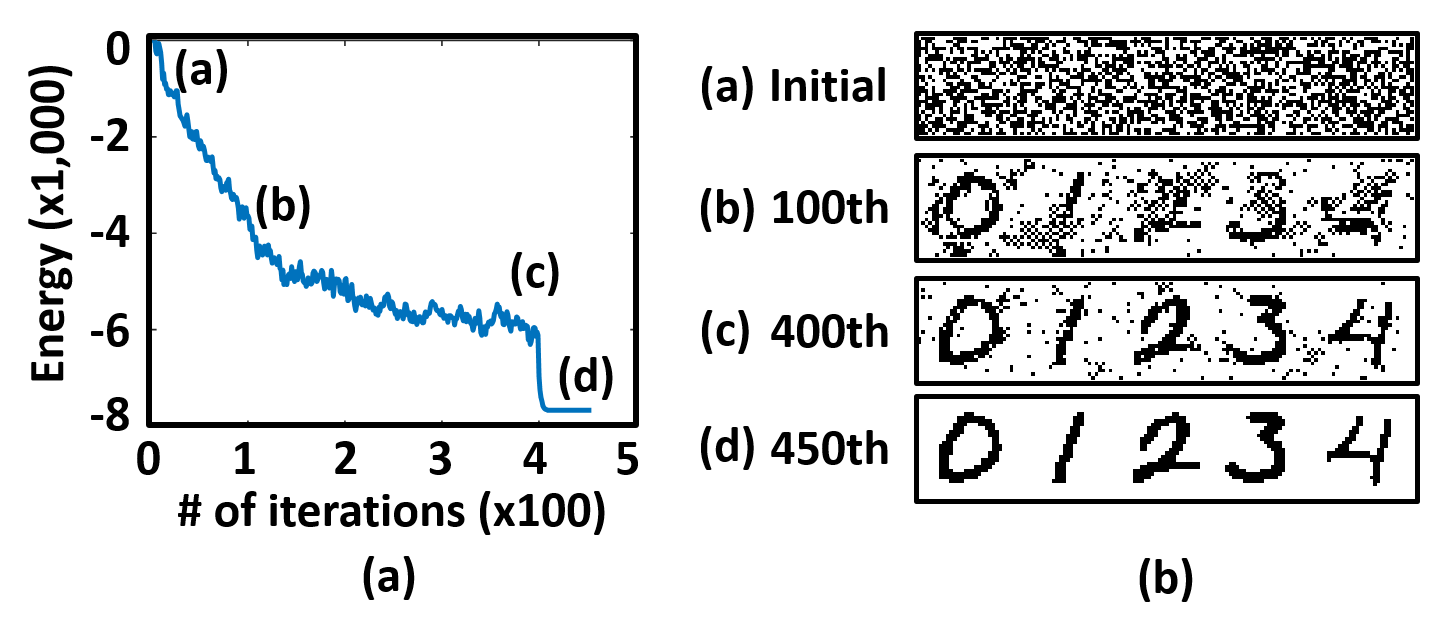}% Here is how to import EPS art
\caption{Application to Maximum-cut problem (a) System energy profile over the number of iterations (spin state update). Based on interconnection weights between spins, the system evolves toward the lowest energy state by updating the spin state. The abrupt energy change at 400 iterations is due to the time-varying switching probability change which can expedite the final process. (b) Visualized spin states at initial, intermediate, and final stages. Initially, spins states are random (with block dots (-1) and white dots (+1)). As more iterations are processed, clear image of digit 0 to 4 is shown.}
\label{fig8}
\end{figure}

The aforementioned generic methodology was the applied to two practical combinatorial optimization problems - Maximum-cut and Graph coloring. First, the proposed Ising spin model is used to solve Maximum-cut problem. 
% definition of the maximum-cut problem
The problem can be defined as finding two mutually exclusive subsets of spins by connecting edges (which connect spins from two separate subsets) so as to maximize the summation of weights along the edges (i.e. boundary between two subsets) \cite{goemn_mc}.
We have used $\sim$3,900 spins for this application and the interconnection weights are programmed so that the spin states show the digit numbers from 0 to 4 without noise once the system reaches the minimum possible energy state. Fig. \ref{fig8} shows the results of the maximum-cut problem and also transition of the system energy over iterations. As shown in Fig. \ref{fig8}(a), the system energy shows a steady drop over the iterations and reaches a minimum after $\sim$450 iterations. The abrupt energy change at the 400th iteration happens due to the time-varying switching probability function as discussed in the previous section. This allows the system to have less random flipping due to the `natural annealing', thereby leading to faster convergence. 
Fig. \ref{fig8}(b) shows visualized spin states at the initial, intermediate and final steps. Here the black dot denotes spin state -1 and white dot represent spin state +1. Initially, spin states start from random distribution of -1 and +1 (black and white dots). As the states are updated through the coupling with neighboring states, target digit numbers are visible with some noise (100th iteration). After 400 iterations, the system almost finds the solution, but, still there exists nontrivial amount of noise mainly due to the ``natural annealing''. The clear output image shows that the optimum solution of the current maximum-cut problem has been obtained.

\begin{figure}
\includegraphics[width=3.4in]{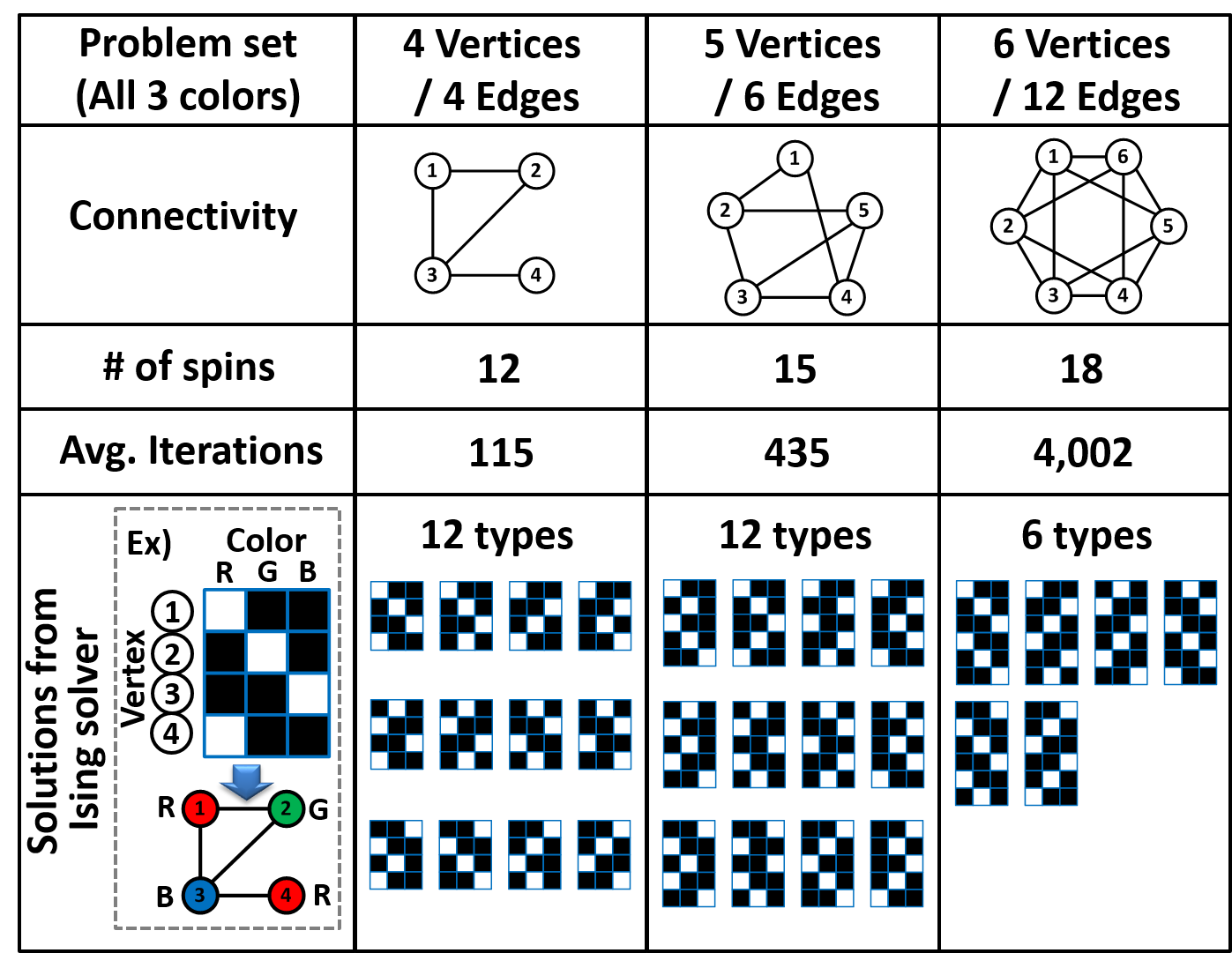}% Here is how to import EPS art
\caption{The proposed Ising spin model is used to solve another combinatorial optimization problem - Graph Coloring. For the problem with n-vertices and k-colors, n $\times$ k spins are needed. Each spin is termed as n$_{i,j}$ where i and j represent current vertex and color respectively. The solutions from the Ising solver can be interpreted as shown inside dotted box on the left. With given 3 types of problems, the Ising model suggests possible solutions.}
\label{fig9}
\end{figure}

Next, our proposed Ising spin model with HM based MTJ was applied to the Graph coloring problem. Graph coloring is a famous NP-complete problem \cite{garey_np} and is defined as ``Is it possible to color n-vertices with k-colors such that no two adjacent vertices have the same color?'' To implement specific hardware to solve the graph coloring problem using our Ising spin model, a pre-processing step is needed to prepare a weight matrix. This can be generated according to the penalty Hamiltonian in \cite{lucas_hamil}. This Hamiltonian basically defines rules to be obeyed and applies an energy penalty whenever these rules are violated. For example, one term of the Hamiltonian for this particular problem provides energy penalty once the edge connects two vertices with the same color. 
Consequently, optimum solution can be obtained when there is no energy penalty, hence the system evolves towards minimum energy state. Interested readers are directed to Ref. \cite{lucas_hamil} for more details on the Hamiltonian for the Ising model. Once the weight matrix is prepared, it shows interconnect map and also weights for each interconnections. We prepared 3 simple Graph coloring problems and implemented corresponding hardware based on weight matrix from penalty Hamiltonian. Fig. \ref{fig9} shows the details of the problem from our proposed Ising solver. 
It is worth noting here that since the spin can only have one of the two states, a total of n $\times$ k spins are needed to represent all possible states for the graph coloring problem (with n-vertices and k-colors). In this case, each spin state is denoted as v$_{i,j}$, where i represents current vertex and j represents current color. For example, v$_{1,1}$ can be interpreted as a spin representing vertex 1 and color 1. The simulations were conducted 1,000 times for each problem to get an average number of iterations to reach minimum energy state. The transition of the system energy is monitored by checking the states of all the spins after each epoch to check if the system has found a solution.

Lastly, let us briefly discuss the energy consumption of the HM based MTJ device used in the Ising spin model. The operation of a single Ising cell can be divided into three parts - read operation, write operation and relaxation time. The time duration for write, relax, and read cycle was taken to be  3 ns, 6 ns, and 1 ns, respectively. The energy consumption during the write cycle is mainly due to the input charge current flowing through the HM layer. Assuming an average input current of $\sim$90 $uA$, the energy consumption during the write operation is evaluated to be $\sim$0.27 $pJ$ with VDD of 1$V$. 
Likewise, the device-circuit simulation of the read circuit yielded an average energy consumption of $\sim$0.04 $pJ$, when considering the average read current to be $\sim$38 $uA$ and VDD to be 1V. In contrast, the energy consumption involved in the relax mode and CMOS switches resulted in insignificant contribution ($\sim$0.01 $pJ$) to the total energy consumption. 
Overall, the proposed single spin model based on our HM based MTJ along with peripheral circuits consumes $\sim$0.32 $pJ$ of energy per single spin update operation.  Even though the write operation consumes most of the energy required ($\sim$83 \%), we believe with improvement in material parameters, write energy can be significantly reduced. 

\section{conclusions}

In summary, we have proposed SHE-MTJ based Ising cell to solve combinatorial optimization problems.
We demonstrate the mapping of annealing and majority vote functions to the behavior of the spins in the nanomagnet. 
Although, the stochastic switching nature of the MTJ due to the thermal noise is usually regarded as a problem in typical memory applications \cite{nigam_sttmram}, such random switching can be exploited to build Ising spin model having the property of ``natural annealing''. Also, the decoupled write and read path through the HM underlayer enables the majority vote function with minimum number of devices. 
Using coupled magnetization dynamics and SPICE simulations, we present solutions for two combinatorial optimization problems - Maximum-cut and Graph coloring.
We believe that the behavior of our HM based MTJ device, mimicking the key elements of the Ising spin model, can potentially pave the way for  scalable, low-power, and simple Ising solver capable of handling complex combinatorial optimization problems.\\

\begin{acknowledgments}
The work is supported in part by Center for Spintronic Materials, Interfaces, and Novel Architectures (C-SPIN), a MARCO and DARPA sponsored StarNet center, the Semiconductor Research Corporation, the National Science Foundation, Intel Corporation, and the National Security Science and Engineering Faculty Fellowship.
\end{acknowledgments}

\bibliography{aps_ising}% Produces the bibliography via BibTeX.

\end{document}